\documentclass[prd,twocolumn,a4paper,superscriptaddress,floatfix]{revtex4}
\usepackage{graphicx}
\usepackage{bbm}
\usepackage[all]{xy}
\usepackage{amsmath}
\usepackage{amssymb}
\usepackage{epstopdf}

\def\be{\begin{equation}}
\def\ee{\end{equation}}
\newcommand{\bq}{\begin{eqnarray}}
\newcommand{\eq}{\end{eqnarray}}
\newcommand{\bes}{\begin{subequations}}
\newcommand{\ees}{\end{subequations}}

\def\ben{\begin{eqnarray}}
\def\een{\end{eqnarray}}
\def\ba{\begin{array}}
\def\ea{\end{array}}

\begin{document}
\newcommand{\half}{{\textstyle\frac{1}{2}}}
\allowdisplaybreaks[3]
\def\a{\alpha}
\def\b{\beta}
\def\g{\gamma}\def\G{\Gamma}
\def\d{\delta}\def\D{\Delta}
\def\ep{\epsilon}
\def\et{\eta}
\def\z{\zeta}
\def\t{\theta}\def\T{\Theta}
\def\l{\lambda}\def\L{\Lambda}
\def\m{\mu}
\def\f{\phi}\def\F{\Phi}
\def\n{\nu}
\def\p{\psi}\def\P{\Psi}
\def\r{\rho}
\def\s{\sigma}\def\S{\Sigma}
\def\ta{\tau}
\def\x{\chi}
\def\o{\omega}\def\O{\Omega}
\def\k{\kappa}
\def\pa {\partial}
\def\ov{\over}
\def\br{\\}
\def\ud{\underline}

\newcommand\lsim{\mathrel{\rlap{\lower4pt\hbox{\hskip1pt$\sim$}}
    \raise1pt\hbox{$<$}}}
\newcommand\gsim{\mathrel{\rlap{\lower4pt\hbox{\hskip1pt$\sim$}}
    \raise1pt\hbox{$>$}}}
\newcommand\esim{\mathrel{\rlap{\raise2pt\hbox{\hskip0pt$\sim$}}
    \lower1pt\hbox{$-$}}}
\newcommand{\dpar}[2]{\frac{\partial #1}{\partial #2}}
\newcommand{\sdp}[2]{\frac{\partial ^2 #1}{\partial #2 ^2}}
\newcommand{\dtot}[2]{\frac{d #1}{d #2}}
\newcommand{\sdt}[2]{\frac{d ^2 #1}{d #2 ^2}}    

\title{$p$-brane dynamics in $(N+1)$-dimensional FRW universes: a unified framework}

\author{L. Sousa}
\email[Electronic address: ]{laragsousa@gmail.com}
\affiliation{Centro de F\'{\i}sica do Porto, Rua do Campo Alegre 687, 4169-007 Porto, Portugal}
\affiliation{Departamento de F\'{\i}sica da Faculdade de Ci\^encias
da Universidade do Porto, Rua do Campo Alegre 687, 4169-007 Porto, Portugal}
\author{P.P. Avelino}
\email[Electronic address: ]{ppavelin@fc.up.pt}
\affiliation{Centro de F\'{\i}sica do Porto, Rua do Campo Alegre 687, 4169-007 Porto, Portugal}
\affiliation{Departamento de F\'{\i}sica da Faculdade de Ci\^encias
da Universidade do Porto, Rua do Campo Alegre 687, 4169-007 Porto, Portugal}

\begin{abstract}

We develop a velocity-dependent one-scale model describing $p$-brane dynamics in flat homogeneous and isotropic backgrounds in a unified framework. We find the  corresponding scaling laws in frictionless and friction dominated regimes considering both expanding and collapsing phases.

\end{abstract} 
\maketitle

\section{Introduction}

Cosmological inflation \cite{Guth:1980zm,Linde:1981mu} is a period of accelerated expansion in the early universe providing the most plausible solution to the flatness, horizon and magnetic monopole problems and explaining the origin of large-scale struture. In the string theory inspired brane inflationary scenario \cite{Dvali:1998pa,Burgess:2001fx,Alexander:2001ks,GarciaBellido:2001ky,Jones:2002cv}, the inflaton may be identified with the distance between two branes or a brane and anti-brane whose interactions originate the inflaton potential. In the latter case (brane-anti-brane inflation) the inflationary epoch ends as a result of brane collision and annihilation and, in the process, lower-dimensional branes are produced by the Kibble mechanism and appear as topological defects. Although the production of $1$-branes (i.e. cosmic strings) is favoured, higher-dimensional $p$-branes may also be generated \cite{Majumdar:2002hy,Sarangi:2002yt}. As a consequence, inflation may offer a natural mechanism for the formation of $p$-brane defect networks evolving in a higher dimensional spacetime.

The large-scale dynamics of cosmic string networks in $(3+1)$-dimensional Friedmann-Robertson-Walker (FRW) backgrounds has been extensively studied using a Velocity-dependent One-Scale (VOS) Model \cite{Martins:1996jp,Martins:1995tg,Martins:2000cs}, which accurately describes the cosmological evolution of the networks' root mean square (RMS) velocity and characteristic length. In \cite{Avgoustidis:2004zt}, this model was generalized to isotropic and anisotropic $(N+1)$-dimensional backgrounds. Furthermore, a similar model  \cite{Avelino:2005kn,Avelino:2010qf,Avelino:2011ev} developed for domain wall networks in isotropic backgrounds was shown to successfully describe the results of high-resolution field theory numerical simulations \cite{Avelino:2008ve}. In this paper, we develop a more general model describing the dynamics of $p$-brane networks of arbitrary dimension in $(N+1)$-dimensional homogenous and isotropic universes in a single framework.

The paper is organized as follows. In Sec. \ref{branedyn} we obtain the $p$-brane equation of motion in a $(N+1)$-dimensional FRW background and apply it to the particular case of cosmic strings. In Sec. \ref{VOS} we derive the VOS equations of motion for $p$-brane networks. In Sec. \ref{frictionless}, we investigate the different frictionless scaling regimes that arise in collapsing and expanding FRW universes and in Sec. \ref{friction} we study the friction dominated regimes. We then conclude  in Sec. \ref{conc}.

\section{$p$-brane dynamics\label{branedyn}}

Consider a $p$-brane whose thickness is much smaller than its curvature radii (zero-thickness limit), so that its world-history may be represented by
\be
x^{\mu}=x^{\mu}(u_{{\tilde \nu}})
\ee
where $u_{\tilde \nu}$ with ${\tilde \nu}=0,1,...,p$ are the coordinates parametrizing the $(p+1)$-dimensional world-volume swept by the $p$-brane, $u_0$ is a timelike parameter and $u_i$ are spacelike parameter. The $p$-brane equation of motion may be obtained by minimizing the action 
\be
S=-\sigma_p \int d^{p+1} u \sqrt{-{\tilde g}}\,,
\ee
with respect to variations of $x^{\mu}$. Here,  ${\tilde g}_{{\tilde \mu} {\tilde \nu}}=g_{\alpha \beta} x^{\alpha}_{,{\tilde \mu}}x^{\beta}_{,{\tilde \nu}}$ and $\sigma_p$ is the $p$-brane mass per unit $p$-dimensional area.

Prior knowledge of the trajectory described by the brane throughout its evolution allow us to define a real scalar field multiplet, $\phi$, in the $p$-brane world-volume, described by the Lagrangian
\be
{\mathcal{L}}=X-V(\phi^e)\,,
\ee 
where $X=-{\phi^e}_{,{\tilde \mu}} \, \partial^{\tilde \mu} \phi^{e,{\tilde \mu}} / 2$, a comma represents a partial derivative, and $V(\phi^e)$ is the potential (on the remainder of this section we shall omit the index $e$). The potential, $V$, needs to have, at least, two degenerate minima in order to admit $p$-brane solutions which can be made arbitrarily thin by appropriate tuning. The initial conditions may be set up in such a way that the scalar field $\phi$ defines a new $p$-brane whose velocity coincides with the velocity of the original $p$-brane. The new $p$-brane may be regarded as a domain wall (a $(p-1)$-brane in the $(p+1)$-dimensional spacetime spanned by the brane during it time evolution), whose dynamics is identical to that of the original $p$-brane. By varying the action,
\be
\mathcal{S}=\int \mathcal{L}\sqrt{-\tilde g}d^{p+1}u,
\ee
with respect to $\phi$, one finds the equation of motion
\be
\frac{1}{\sqrt{-\tilde g}}\left(\sqrt{-{\tilde g}}\phi^{,\bar \mu} \right)_{,\bar \mu}=-V_{,\phi}\,.
\label{eom_phiw1}
\ee

Consider a flat FRW universe whose line element is given by
\be
ds^2=a^2(\eta)(-d\eta^2+{\bf dx} \cdot {\bf dx})\,,
\label{frw}
\ee 
where $a(\eta)$ is the scale factor, $\eta=dt/a$ is the conformal time, $t$ is the physical time and ${\bf x}$ are 
comoving coordinates.

One may identify the timelike coordinate with the conformal time ($u_0=\eta$), so that ${\tilde g}_{00}=a^2(\eta)$. Moreover, the velocity of the brane may be taken to be orthogonal to the brane itself and, therefore, perpendicular to all spatial parameters of the brane (temporal-transverse gauge conditions)
\be
{\dot {\bf x}} \cdot  {\bf x}_{,u_{\tilde i}}=0 \Rightarrow {\tilde g}_{0{\tilde i}}={\tilde g}_{{\tilde i}0}=0\qquad  {\tilde i}=1,...,p+1\,.
\ee
Consider \textbf{a} set of local spatial coordinates $(u_1,...,u_p, u_{p+1})$ such that the brane is locally a coordinate surface for which $u_{p+1}$ is constant and \textbf{it} moves along this direction. The spatial coordinates $(u_1,...,u_p)$ may be chosen in such a way that they form an orthogonal set which locally parametrizes the brane and, for latter convenience, whose coordinate lines coincide with its principal directions of curvature. Since the $p$-brane is embedded in a flat FRW Universe, this coordinate system exists in the vicinity of any non-umbilic point of brane's surface \cite{topogonov}. 
By introducing the scale factors of the coordinate system $|{\bf x}_{,u_{\tilde i}}|=h_{\tilde i}$, the metric components in this coordinate system can be written as
\begin{equation}
{\tilde g}_{{\tilde i}{\tilde j}}= \left\{
\begin{array}{cl}
a^2h_{\tilde i}^2\,, \quad & \mbox{if ${\tilde i}={\tilde j}$}\,, \\
0\,, \quad & \mbox{if ${\tilde j}\neq{\tilde i}$}\,,
\end{array}
\right.
\end{equation}
so that ${\tilde g}=-a^{2(p+2)}h_1^2 ... h_{p+1}^2$. Eq. (\ref{eom_phiw1}) then becomes
\be
\ddot{\phi}+p{\mathcal H} \dot{\phi}-\nabla^2_{\bf u}\phi=-a^2 V_{,\phi}\,,
\label{eom_phiw2}
\ee
with ${\mathcal H}={\dot a}/a$ and where

\begin{equation}
\nabla^2_{\bf u}\phi = \left(\prod_{j=1}^{p} h_j\right)^{-1}\left(\frac{\partial}{\partial u_{p+1}}\left(\frac{\partial\phi}{\partial u_{p+1}} \prod_{j=1}^{p} h_j\right)\right)
\end{equation}
is the laplacian for this set of coordinates. Here, we considered the zero-thickness limit, neglecting the variation of the scalar field $\phi$ on the directions tangent to the brane \cite{Garfinkle:1989mv} ($\partial \phi/\partial u_i=0$ for $i=1,...,p$). We also  taken $h_{p+1}=1$ so that $du_{p+1}$ is the infinitesimal arclength along the $u_{p+1}$ direction.

Given Eq. (\ref{eom_phiw2}) and using the method described in detail in \cite{Sousa:2010zz}, one obtains the equation of motion
\be
{\dot v}+\left(1-v^2\right)\left[\left(p+1\right)\mathcal{H}v-\kappa \right]=0\,,
\label{eom_v}
\ee
where
\be
\kappa=\hat{\bf{v}}\cdot \sum_{{\tilde i}=1}^{p} {\bf k}_{\tilde i}\,,
\label{dyng}
\ee
and ${\bf k}_{\tilde i}$ is the comoving curvature vector along the principal direction of curvature $u_i$ and ${\bf v}=v \hat{\bf{v}}$ is the brane velocity. Note that $\kappa$ is obtained by projecting the curvature vectors along the velocity direction. In doing so one is  losing information about the acceleration of the brane perpendicular to ${\bf v}$ (normal acceleration) but the evolution of $v$ only depends on the tangential acceleration of the brane.

\subsection{Cosmic strings \label{strings}}

The world history of an infinitely thin cosmic string in a flat FRW universe can be represented by a two-dimensional world-sheet with ${\bf x}={\bf x} (\eta,\sigma)$, obeying the usual Goto-Nambu action. The equations of motion can be written as
\bq
\ddot{{\bf{x}}}+2\mathcal{H}\left(1-\dot{{\bf{x}}}^2\right)\dot{{\bf{x}}} &=& \epsilon^{-1}\left(\epsilon^{-1}\,
{\bf{x}}'\right)'\label{vos}\\
\dot{\epsilon} &=& -2\mathcal{H}\epsilon \dot{{\bf{x}}}^2\,,
\eq
with
\be
\dot{\bf{x}} \cdot {\bf{x}}'  =  0\,, \quad \quad \epsilon  = \left(\frac{{\bf{x}}'^2}{1-\dot{{\bf{x}}}^2}\right)^{\frac{1}{2}}\,, \label{epsilon}
\ee
where dots and primes are derivatives with respect to $\eta$ and $\sigma$, respectively. Consider
\be
\hat{\bf{v}}=\frac{\dot{\bf{x}}}{v}\,,\quad \quad \hat{\bf{y}}=\frac{{\bf{x}}'}{|{\bf x}'|}\,,
\ee
where $v(\eta,\sigma)=|\dot {\bf x}|$. The left hand side of Eq.  (\ref{vos}) is given by
\be
\ddot{{\bf{x}}}+2\mathcal{H}\left(1-\dot{{\bf{x}}}^2\right)\dot{{\bf{x}}}=\dot{v}\hat{\bf{v}}+v{\dot {\hat{\bf{v}}}}+2\mathcal{H}\left(1-v^2\right)v\hat{\bf{v}}\,,
\ee
where ${\dot {\hat{\bf{v}}}}$ is perpendicular to $\hat{\bf{v}}$. Moreover, the right hand side of Eq.  (\ref{vos}) gives
\bq
\epsilon^{-1}\left(\epsilon^{-1}{\bf{x}}'\right)' &=& \frac{1}{\gamma}\frac{\partial}{\partial l}\left(\frac{\hat{\bf{y}}}{\gamma}\right)\\\nonumber &=&\left(\frac{1}{\gamma^2}{\bf k} -v \frac{\partial v}{\partial l}\hat{\bf{y}}\right)\,,
\eq
where we have taken into account that $\gamma=(1-v^2)^{-1/2}$, $\partial \hat{\bf{y}}/\partial l={\bf k} \perp \hat{\bf{y}}$ and the fact that the physical length along the string is given by $dl=|d {\bf x}| =  {|{\bf x}'|} d\sigma$. Henceforth, the component of Eq. (\ref{vos}) parallel to ${\bf v}$ yields
\be
{\dot v}+(1-v^2)\left(2\mathcal{H}v-\kappa\right)=0\,,
\ee
where $\kappa={\bf k} \cdot \hat{\bf{v}}$, which is equivalent to Eq. (\ref{eom_v}) in the particular case with $p=1$.

It is interesting to contrast this with the corresponding result for domain walls, in which case the coefficient of the cosmological damping term is $N$ (rather than $2$) and $\kappa=\hat{\bf{v}}\cdot \sum_{{\tilde i}=1}^{N-1} {\bf k}_{\tilde i}$ (rather than $\hat{\bf{v}} \cdot {\bf k}$). Hence, the differences between the macroscopic evolution of cosmic string and domain wall networks 
are expected to increase with $N$.

\section{VOS model\label{VOS}}

Let us now consider the case of a network of $p$-branes in a $(N+1)$-dimensional FRW Universe. The RMS velocity, denoted by ${\bar v} ={\sqrt {\langle v^2\rangle}}$, is defined as
\be
{\bar v}^2 = \frac{\int v^2 \rho dV}{\int \rho dV}\,,
\ee
where $\rho$ is the $p$-brane energy density and $V$ is the physical volume. An equivalent definition for 1-branes would be
${\bar v}^2=\int v^2 \gamma dl/\int \gamma dl$. The characteristic length, $L$, of the network is defined as
\be
{\bar \rho} = \frac{\sigma_p}{L^{N-p}}\,,
\ee
where ${\bar \rho} = V^{-1}\int \rho dV$ is the average brane density. An alternative but less useful definition of the physical lengthscale of the network would be
\be
\langle \rho/\gamma \rangle=\frac{\sigma_p}{L_{\rm ph} ^{N-p}}\,,
\ee
where $\langle...\rangle$ denotes a volume weighted average.
This definition has the advantage that $L_{\rm ph}$ is sensitive only to the spatial profile of the network, and is therefore independent of the velocity distribution. However, since a direct relation between $L_{\rm ph}$, ${\bar \rho}$ and ${\bar v}$ does not exist, $L$ is often a more useful variable than $L_{\rm ph}$. In any case, $L$ and $L_{\rm ph}$ are in general very similiar, except if the $p$-branes are ultra-relativistic.

Multiplying Eq. (\ref{eom_v}) by $v$, making the volume average and then dividing by ${\bar v}$, one obtains
\be
{\dot {\bar v}}+\left(1-{\bar v}^2\right)\left[\left(p+1\right)\mathcal{H}{\bar v}-{\bar \kappa} \right]=0\,,
\label{VOSk1}
\ee
where 
\be
{\bar \kappa}=\frac{\left<v\left(1-v^2\right) \kappa \right>}{{\bar v}\left(1-{\bar v}^2\right)}=\frac{\int v\left(1-v^2\right)\kappa\rho dV}{{\bar v}\left(1-v^2\right)\int\rho dV}
\ee
and we have made the assumption that $\left<v^4\right>={\bar v}^4$ (see \cite{Martins:1996jp}). Eq. (\ref{VOSk1}) may also be written as
\be
\frac{d{\bar v}}{dt}+\left(1-{\bar v}^2\right)\left[\frac{{\bar v}}{\ell_d}-\frac{k}{L}\right]=0\,,
\label{VOS_v}
\ee
where $\ell_d^{-1}=(p+1)H$ is the damping lengthscale, $H={\mathcal H}a$ is the Hubble parameter, $k={\bar \kappa}L/a$ is a dimensionless curvature parameter equivalent (for $p=1$) to that of the original VOS model for cosmic strings  \cite{Martins:1996jp,Avgoustidis:2004zt}. The frictional force originated by the interaction of the branes with ultrarelativistic particles may be included in Eq. (\ref{VOS_v}), by introducing an extra term in the damping lenghtscale, $\ell_d^{-1}=(p+1)H+\ell_f^{-1}$. The friction lengthscale, $\ell_f$, will be properly defined in in Sec. \ref{friction}.

Energy-momentum conservation in a FRW universe then implies that
\be
\frac{d{\bar \rho}}{dt}+NH\left({\bar \rho}+{\bar {\mathcal P}}\right)=0\,,
\label{em-frw}
\ee
assuming that the $p$-brane network is statistically homogenous and isotropic on large scales. Here, ${\bar {\mathcal P}}=V^{-1}\int\mathcal{P}dV$ is the average brane pressure. The equation of state parameter of the brane gas is given by \cite{Boehm:2002bm}:
\be
w=\frac{\bar {\mathcal P}}{\bar \rho}=\frac{1}{N}\left[\left(p+1\right){\bar v}^2-p\right]\,.
\label{eos}
\ee
We may also include an extra term in Eqs. (\ref{eos}) and (\ref{em-frw})  to account for the energy loss due to $p$-brane collapse and friction
\be
\frac{d{\bar \rho}}{dt}+H{\bar \rho}\left[D+\left(p+1\right){\bar v}^2\right]=-\left(\frac {{\tilde c}{\bar v}}{L}+\frac{{\bar v}^2}{\ell_f}\right){\bar \rho}\,,
\label{vos-den}
\ee
where ${\tilde c} \ge 0$ is the energy-loss parameter and $D=N-p$.
Consequently, the equation of motion for the characteristic lengthscale of the network yields
\be
\frac{dL}{dt}=HL+\frac{L}{D \ell_d}{\bar v}^2+ \frac{{\tilde c}}{D}{\bar v}\,.
\label{vos-L}
\ee
Eqs. (\ref{VOS_v}) and (\ref{vos-L}) constitute a unified VOS model for the dynamics of $p$-brane networks in $(N+1)$-dimensional FRW universes. In the following section, we shall obtain the corresponding scaling laws. The reader is referred to \cite{Avelino:2011ev} for a more detailed discussion of the more specific domain wall scenario with $p=N-1$.

\section{Frictionless regimes \label{frictionless}}

Let us assume that the dynamics of the universe is driven by a fluid with $w={\rm constant} \neq -1$ so that $a \propto t_*^\beta$, where $\beta=2/(N(w+1))$ and $t_* \ge 0$ is the time elapsed since the initial singularity (if $dt_*=dt$) or the time remaining up to the final singularity (if $dt_*=-dt$) at $t_*=0$. We shall consider six different models labeled by $M^s_i$, where $s=\pm$ depending on whether $dt=\pm dt_*$ and $i=1,2$ or $3$,  depending on whether $\beta <0$, $0<\beta <1$ or 
$\beta >1$, respectively. The models $M^+_2$, $M^+_3$ and $M^-_1$ represent expanding 
solutions with $t_*=0$ either at the the big-bang ($M^+_2$ and $M^+_3$) or at the big rip (for $M^-_1$). The models $M^+_1$, $M^-_2$ and $M^-_3$ represent collapsing  universes with $t_*=0$ either at the the big-crunch ($M^-_2$ and $M^-_3$) or at an  initial infinite density singularity with $a_*=\infty$ (for $M^+_1$).

\subsection{Linear scaling solutions ($\ell_f=\infty$)}

If the friction lengthscale becomes negligible compared to the Hubble radius then Eqs. (\ref{VOS_v}) and (\ref{vos-L}) may have a linear attractor solution. This attractor solution corresponds to a linear scaling regime of the form
\be
L=\xi t_* \qquad \mbox{and} \qquad \bar v=\mbox{constant}\,,
\label{linearscalinga}
\ee
with 
\be
\xi = \sqrt{\left|\frac{k(k+{\tilde c})}{\beta (1-\beta)D(p+1)}\right|} \quad{\bar v} = \sqrt{\frac{(1-\beta)kD}{\beta(k+{\tilde c})(p+1)}}\,. \label{linearscaling}
\ee
The conditions  $0 < {\bar v} < 1$ and $\xi > 0$ are sufficient to show that models $M_3^+$ and $M_1^-$  do not admit linear scaling solutions, which would require a negative energy loss parameter ${\tilde c}$.

The RMS velocity, ${\bar v}$, of maximally symmetric $p$-branes with a $S_{p-i}\otimes \mathbbm{R}^i$ topology oscillating periodically in a Minkowski spacetime is given by 
\be
v_{\rm min} \le {\sqrt {\frac{p-i}{p-i+1}}} \le v_{\rm max} \label{averagevel}\,,
\ee
with $v_{\rm min}^2=1/2$ and $v^2_{\rm max}=p/(p+1)$, which correspond, respectively, to $i=p-1$ (only one of the principal curvatures is non-zero) and $i=0$ (for fully spherically symmetrical $p$-branes). For $\beta=0$, the curvature parameter $k$ must be equal to zero for a linear scaling solution with ${\bar v} \le 1$ to be attained. The expansion (collapse) of the universe hinders (aids) the velocity of the branes, leading to a  smaller (larger) RMS velocity and a curvature parameter, $k$, larger (smaller) than zero. Therefore, one expects that
\bq
0 & < & {\bar v}<v_{\rm max}, \quad \mbox{for $M_2^+$}\,,\\ \nonumber
v_{\rm min} & < & {\bar v}<1, \quad \mbox{for $M_1^+$ and $M_3^-$}\,.
\eq

On the other hand the characteristic lengthscale of the network is necessarily constrained by causality and, as a consequence, $L$ is expected to be bounded by the particle horizon at any given time. In the case of models $M_1^+$, $M_2^+$ and $M_3^-$ this implies that
\be
L < d_H = \int_{t_i}^t \frac{dt'}{a(t')} = \frac{t_*}{\left|1-\beta\right|}\label{causality}\,,
\ee
with $t_i=0$ or $t_i=-\infty$ (depending on whether $s=+$ or $-$, respectively), or equivalently 
\be
{\bar v}^2 < (k+{\tilde c})^{-2}\label{vcons3}\,.
\ee

Some remarks regarding the energy-loss term are necessary. For simplicity, we shall consider the most trivial case of flat 
$p$-branes. A moving $p$-brane spans a $q$-dimensional surface (with $q=p+1$) which has $N-q$ degrees of freedom. Hence, if 
$N \le 2(N-q)$ then two flat $q$-dimensional surfaces do in general intersect but that is no longer true if $N > 2(N-q)$ (or equivalently $p<(N-1)/2$). Hence, if the $p$-branes are thin ($p$-brane thickness much smaller than $L$), the energy-loss parameter ${\tilde c}$ may be much smaller than unity if $p<(N-1)/2$. Note, however, that the linear scaling solution in Eq. (\ref{linearscalinga}) may be attainable for $\beta > 1-p/N$, even if ${\tilde c}=0$ with 
\be
\xi = \sqrt{\left|\frac{k^2}{\beta (1-\beta)D(p+1)}\right|} \quad{\bar v} = \sqrt{\frac{(1-\beta)D}{\beta(p+1)}}\,. \label{linearscalingc0}
\ee
As a consequence, even if ${\tilde c} \ll 1$, the $p$-brane network may be able to attain a linear scaling regime in the $M_2^+$ model. However, if ${\tilde c}=0$, a linear scaling solution is no longer possible in a collapsing universe.

\subsection{Inflation and superinflation}

In the case of models $M_3^+$ and $M_1^-$ the expansion is fast enough ($\ddot{a}>0$) to decelerate the branes and make the RMS velocity arbitrarily small. As a consequence, the inflationary models $M_3^+$ and $M_1^-$ undergo a stretching regime described by the scaling laws
\be
L \propto a \,, \qquad v \propto (Ha)^{-1} \propto a^{-1-1/\beta} \to 0\,.
\ee

If the $p$-branes are the dominant energy component of the universe then
\be
\beta =\frac{2}{N(1+w_b)}= \frac{2}{D+(p+1) {\bar v}^2}\,.
\ee
In order to accelerate the universe one needs $\beta >1$ (or equivalently $w_b <w_c= (2-N)/N$) and, consequently
\be
{\bar v}^2 < \frac{2-D}{p+1}\,.
\ee 
We may then conclude that only domain walls ($D=N-p=1$) are able to drive an inflationary  phase.

\subsection{Ultra-relativistic collapsing solution}

In the case of model $M^-_2$ (which represents a collapsing universe with $k<0$ and $0<\beta <1$), the comoving Hubble radius, $|H^{-1}|$ decreases with time. As a consequence, the curvature scale of the $p$-brane will necessarily become smaller than $|H^{-1}|$ and will tend to be conformally contracted in physical coordinates whilst travelling at ultrarelativistic speeds with
\be
L_{\rm ph} \propto a \,, \qquad {\bar \gamma} \propto a^{-(p+1)}\,,
\ee
where is defined by ${\bar \gamma} =(1-{\bar v}^2)^{-1/2}$ and $L \sim {\bar \gamma}^{1/D} L_{\rm ph} \propto  a^{-(N+1)/D}$. Hence, at ultra-relativistic speeds $L$ ceases to be an accurate measure of $L_{\rm ph}$.

\section{Friction-dominated regimes\label{friction}}

The interaction of $p$-branes with the ultrarelativistic particles in a radiation fluid results in a frictional damping of the form
\be
\frac{d {\mathbf v}}{dt}=-\frac{1}{\ell_f} (1-v^2) \mathbf{v}
\ee
where $\ell_f\propto \sigma_p \lambda^{p+1-N}/\rho_ {\rm rad} \propto a^{p+2}$ is the friction lengthscale, $\lambda \propto a$ and $\rho_{\rm rad}\propto a^{-(N+1)}$ are, respectively, the typical wavelength and the energy density of relativistic particles.

\subsection{Expanding universe}

If, at the moment of formation, the density of $p$-branes is sufficiently low ($HL \gg {\bar v}^2 L/\ell_f$ and $HL \gg {\tilde c}{\bar v}$), the $p$-brane network will experience a stretching regime. During this regime, the network will be conformally 
stretched with 

\be
L \propto a \quad \mbox{and}\quad {\bar v}\propto \ell_f/a \, .
\ee 
As ${\bar v}$ increases, the network will start to experience a considerable energy loss due to self-interaction ($HL\sim{\tilde c}{\bar v}$) and, as a consequence, the Kibble regime, caracterized by the scaling laws
\be
L\propto {\sqrt{\ell_f/H}}\quad \mbox{and} \quad {\bar v} \propto{\sqrt{\ell_f H}} \, ,
\ee
 emerges. If, at the moment of formation of the branes, the density of the network is high enough, it will not experience the stretching regime and the Kibble regime may occur right away. The friction lengthscale, $\ell_f$, grows at a higher rate than $H^{-1}$ (for $p+1>1/\beta$) as the universe expands 
and eventually overcomes the characteristic length, $L$. This implies that the Kibble regime is necessarily transient.

\subsection{Collapsing universe}

In the case of a collapsing universe, the evolution of the network ends in a friction dominated era, with the $p$-branes coming to a standstill in comoving coordinates and then being conformally contracted with ${\bar \rho}\propto a^{-1}$. The background temperature and density will eventually approach those of the brane-forming phase transition and, as a consequence, the branes will dissolve into the high density background.

\section{Conclusions \label{conc}}

In this paper we developed a VOS model describing the dynamics of $p$-brane networks in homogeneous and isotropic backgrounds. We used it to determine the evolution of the networks' characteristic length and RMS velocity in flat expanding or collapsing FRW Universes, obtaining the corresponding scaling laws in frictionless and friction dominated regimes.
Our model provides a unified semi-analitic description of $p$-brane network dynamics, highlighting the common and 
distinct features characterizing the evolution of $p$-brane networks of different dimensionality. The connection between cosmic string and domain wall network evolution in $3+1$ dimensions, which is clearly established in this work, is of particular relevance. Although we have assumed throughout the paper that all $p$-branes have the same tension, our model may be extended to incorporate the possibility of multi-tension networks. 




\bibliography{pVOS}

\end{document}